 \def\Title{``Quantal'' behavior in classical probability}
 \def\arXiv{quant-ph/0106072 v6}
\def\Abstract{%
A number of phenomena generally believed characteristic of quantum mechanics and seen as
interpretively problematic---the incompatibility and value-indeterminacy of variables,
the non-existence of dispersion-free states, the failure of the standard
marginal-probability formula, the failure of the distributive law of disjunction and
interference---are exemplified in an emphatically non-quantal system: a deck of playing
cards. Thus the appearance, in quantum mechanics, of incompatibility and these
associated phenomena requires neither explanation nor interpretation.
}%
 \documentclass[aps,nobibnotes,nofootinbib,twoside,noshowpacs,draft]{revtex4}%
 \usepackage[tbtags,sumlimits]{amsmath}
 \usepackage{amssymb}
 \makeatletter
 \renewcommand\NAT@citesuper[3]{%
  \ifNAT@swa
   \leavevmode
   \unskip
   \textsuperscript{(#1)}
   \if*#3*\else\ (#3)\fi
  \else
   #1%
  \fi
  \endgroup
 }%
 \makeatother

 \bibpunct{}{}{,}{s}{}{}

\makeatletter

\def\p@section{}

\def\p@subsection{}

\def\p@subsubsection{}

\def\p@paragraph{}

\def\p@subparagraph{}
 \def\@onlinecite#1{\begingroup\let\@cite\NAT@citenum\citealp{#1}\endgroup}
\makeatother

 \usepackage{xspace}

 \newcommand{\ie}{i.e., }
 \newcommand{\eg}{e.g., }
 
 \newcommand{\Eg}{E.g., }
 \newcommand{\cf}{cf.\xspace}

 \newcommand{\vN}{von Neumann\xspace}
 
 \newcommand{\Schrodinger}{Schr\"odinger\xspace}

 \newcommand{\RefSec}[1]{Sec.~\ref{#1}}

 \newcommand{\RefEqn}[1]{Eq.~\eqref{#1}}
 \newcommand{\RefEqns}[1]{Eqs.~\eqref{#1}}

 \newcommand{\RefCite}[1]{Ref.~\onlinecite{#1}}

 \newcommand{\QED}{\hfill\ensuremath{\square}\smallskip\smallskip}
 \newcommand{\IFF}{\/i\/f\/f\/\xspace}
 \newcommand{\DefEq}{\equiv}
 \newcommand{\abs}[1]{\ensuremath{\left\vert#1\right\vert}}
 \newcommand{\Sum}[1]{\ensuremath{\sum_{#1}}}
 \newcommand{\KDelta}[2]{\ensuremath{\delta_{{#1}{#2}}}}
 \newcommand{\set}[1]{\ensuremath{{\left\{\,#1\,\right\}}}}%

 \newcommand{\Or}{\ensuremath{\,\mathop{\vee}}\,}
 \newcommand{\Orj}[2]{\ensuremath{\mathop{\textstyle\bigvee}_{\!#1}\,{#2}}}

 \newcommand{\Op}[1]{\ensuremath{\mathbf{#1}}\xspace}
 \newcommand{\OpAdj}[1]{\Op{#1^{\dagger}}}

 \newcommand{\andthen}{\ensuremath{\mathop{\&}}}
 \newcommand{\Not}{\ensuremath{\,{\sim}}}
 \newcommand{\suchthat}{\ni}

 \newcommand{\Sys}[1]{\ensuremath{\mathcal{#1}}\xspace}

 \newcommand{\Trace}[2][]{\ensuremath{{\rm Tr}%
                    {\!}_{\Sys{#1}}\Bigl\{\,{#2}\Bigr\}}}

 \newcommand{\ket}[1]{\ensuremath{\vert\,{#1}\,\rangle}}
 \newcommand{\bra}[1]{\ensuremath{\langle\,#1\,\vert}}
 \newcommand{\braket}[2]{\ensuremath{\langle#1\,\vert\,#2\rangle}}
 \newcommand{\proj}[1]{\ensuremath{\ket{{#1}}\bra{{#1}}}}

 \newcommand{\ProjSym}{\boldsymbol{\mathsf{P}}}
 \newcommand{\Proj}[2][]{\ensuremath{\ProjSym^{\Sys{#1}}}[\,#2\,]}

 \newcommand{\bRho}{\pmb{\rho}}

 \newcommand{\Rho}[2]{\ensuremath{\mathop{\bRho
     \def\xx{#1}\ifx\xx\empty{_{#2}}\else{^{\Sys{#1}}_{#2}}\fi}}\xspace}

 \newcommand{\sz}{s^{[0]}}

 \renewcommand{\Pr}[2][]{\ensuremath{{\rm Pr}_{#1}\bigl(\,{#2}\,\bigr)}}
 \newcommand{\Prob}[3][]{\ensuremath{{\rm Pr}_{#1}\bigl(\,{#2}\bigm|#3\,\bigr)}}

 \newcommand{\K}{\ensuremath{\text{\textsf{K}}}\xspace}
 \newcommand{\Q}{\ensuremath{\text{\textsf{Q}}}\xspace}
 \newcommand{\J}{\ensuremath{\text{\textsf{J}}}\xspace}
 
 \renewcommand{\S}{\ensuremath{\text{\textsf{S}}}\xspace}
 \renewcommand{\H}{\ensuremath{\text{\textsf{H}}}\xspace}
 \newcommand{\D}{\ensuremath{\text{\textsf{D}}}\xspace}

 \newcommand{\R}{\ensuremath{\text{\textsf{R}}}\xspace}
 \newcommand{\B}{\ensuremath{\text{\textsf{B}}}\xspace}

 \newcommand{\Face}{\ensuremath{\text{\textsl{Face}}}\xspace}
 \newcommand{\Suit}{\ensuremath{\text{\textsl{Suit}}}\xspace}
 \newcommand{\Color}{\ensuremath{\text{\textsl{Color}}}\xspace}
 \newcommand{\Letter}{\ensuremath{\text{\textsl{Letter}}}\xspace}

 \newcommand{\SysS}{\Sys{S}}
 \newcommand{\SysSvd}{\ensuremath{\SysS_{\text{vd}}}\xspace}
 \newcommand{\SysSvi}{\ensuremath{\SysS_{\text{vi}}}\xspace}

 \newcommand{\RefTable}[1]{Table~\ref{#1}}
 \newcommand{\RefTables}[1]{Tables~\ref{#1}}

 \newcommand{\strt}{\rule[-6pt]{0pt}{18pt}}

 \newcommand{\incompP}{\ref{incompP}\xspace}
 \newcommand{\nojointP}{\ref{nojointP}\xspace}
 \newcommand{\nosharpP}{\ref{nosharpP}\xspace}
 \newcommand{\margP}{\ref{margP}\xspace}
 \newcommand{\interfereP}{\ref{interfereP}\xspace}
 \newcommand{\indetP}{\ref{indetP}\xspace}
 \newcommand{\theseP}{\incompP--\interfereP}
 \newcommand{\allP}{\incompP--\indetP}
 \newcommand{\BohrP}{\incompP--\nosharpP}
 \newcommand{\statP}{\nojointP--\interfereP\xspace}

 \newcommand{\p}[1]{\ensuremath{p_{#1}}\xspace}
 \newcommand{\q}[1]{\ensuremath{q_{#1}}\xspace}

 \newcommand{\EvP}{\ensuremath{E_P}\xspace}

\begin{document}
 \makeatletter
 \def\ps@titlepage{%
   \renewcommand{\@oddfoot}{}%
   \renewcommand{\@evenfoot}{}%
   \renewcommand{\@oddhead}{\hfill\arXiv}
   \renewcommand{\@evenhead}{}}
 \makeatother

\title[\Title] 
      {\Title} 

\author{K.~A.~Kirkpatrick}
\email[E-mail: ]{kirkpatrick@physics.nmhu.edu}
\affiliation{New Mexico Highlands
University, Las Vegas, New Mexico 87701}
\begin{abstract}
 \Abstract
\end{abstract}

 \maketitle
 \makeatletter\markboth{\hfill\@shorttitle\hfill}{\hfill\@shorttitle\hfill}\makeatother
 \pagestyle{myheadings}

\section{Introduction}%
I will show you a probabilistic system which exhibits these phenomena:
\begin{enumerate}\renewcommand{\theenumi}{Q\arabic{enumi}}
\item Observations of the several variables of the system cannot be made
simultaneously---the processes for their observation are mutually
inconsistent.\label{incompP}

\item Variables are \emph{incompatible}: The statistics of the observation of two
different variables in succession depend on the order of their observation; joint
probability distributions of such incompatible variables do not exist.\label{nojointP}

\item The system has no dispersion-free states---if, in a particular preparation of the
system, one variable is sharp, the variable(s) incompatible with it cannot
be.\label{nosharpP}

\item An observation whose result is ignored may affect the statistics of a succeeding,
incompatible, observation---an apparent contradiction of the formula for marginal
probabilities.\label{margP}

\item Under certain circumstances, a disjunction of several values of a variable may
fail to distribute through the conjunction with a succeeding, incompatible,
observation---\emph{interference} may occur.\label{interfereP}
\newcounter{holdenumi}\setcounter{holdenumi}{\value{enumi}}
\end{enumerate}
Would you not assume any system such as this to be a quantum-mechanical one? Such an
assumption would be most reasonable: each phenomenon on this list has been considered, by
one author or another, to be characteristic of quantum mechanics; each has been
considered inexplicable by (and unacceptable from the viewpoint of) classical physics;
each has inspired interpretations of quantum mechanics (Copenhagenism, quantum
probability, quantum logic, \dots; for an excellent review, see \RefCite{Jammer74}). But
that assumption, however reasonable, would be incorrect---the system is unarguably
classical, consisting of playing cards being drawn from a deck under an unusual, but
straightforward, scheme.%
\footnote{The cards are selected using ordinary (chaotic) mechanical shuffling; the
resulting system is deterministic (although probabilistic) and completely describable by
classical mechanics---as far from quantum mechanical as is possible for a physical
system to be.} %

I will argue, in \RefSec{S:Conclusion}, that the existence of such an example
invalidates every call for the interpretation of quantum mechanics, every claim of
metaphysical difficulty regarding quantum mechanics, which is based on the statistical
phenomena \statP. If a problem of meaning or understanding were raised by these
phenomena, it would not be a problem for the understanding of quantum mechanics
\emph{per se} but a problem for the understanding of the category \emph{probabilistic
models of sequences of variable-evaluation events in systems having several
variables}---one such system being quantum mechanics, and another being our example.

In \RefSec{S:fundamentals}, I present the general probabilistic concepts and some
necessary notation. In \RefSec{S:Compatibility}, I give formal definitions of {\statP}
(\RefEqns{E:Compatibility}, \eqref{E:NoSharp}, \eqref{E:MargProbFail}, and
\eqref{E:IntDef}, respectively), and prove that \RefEqn{E:Compatibility}, the definition
of compatibility, is equivalent to the quantum-mechanical definition in terms of
commuting operators. In \RefSec{S:Mimick}, I present examples of a classical
probabilistic system which exhibits each of the properties \theseP; the statistical
properties of these systems are summarized in \RefEqn{E:results} and the discussion
following. In \RefSec{S:Manifestation}, I clarify how \ref{margP} and \ref{interfereP}
may occur (both in classical probability and in quantum mechanics) without violating
basic probability identities: I deal with the problem of the completeness of the values
of variables in the derivation of the marginal-probability formula by introducing the
concept of \emph{manifestation}, and then use this to solve a problem (noted by
Margenau\cite{Margenau63a}) in which the marginal-probability formula seems to fail in
quantum mechanics. In \RefSec{S:Indeterminate}, I discuss the matter of
value-indeterminate variables, called ``nonreality'' in quantum mechanics, and show that
it is a natural occurrence in nondeterministic (as contrasted with simply chaotically
probabilistic) systems. I provide several appendices: Appendix \ref{S:ElemProb} gives a
brief summary of the probability of propositions, Appendix~\ref{S:development} contains
the mathematical analysis of the example system, and Appendix~\ref{S:exercises} contains
several exercises which illustrate \ref{incompP}, \ref{nojointP}, and \ref{margP} in a
very simple system.

\section{The general setting; notation}\label{S:fundamentals}%
In this section I discuss the theory of sequences of events in a probabilistic system
described by more than one variable. This does not require an extension of standard
probability theory (which is summarized in Appendix \ref{S:ElemProb})---merely, to avoid
ambiguities, the introduction of several new terms and some notation (which are new only
because textbooks do not consider systems of several independent variables).

\paragraph*{The system.}
A probabilistic system having several variables, $P$, $Q$, \dots, whose possible values
are discrete: \set{\p{1},\,\p{2},\dots}, \set{\q{1},\,\q{2},\dots}, \dots. The values of
each variable are disjoint and complete (\cf Appendix \ref{S:ElemProb}). (If it doesn't
lead to ambiguity, the proposition that a variable has a particular value will be
abbreviated to the value itself: $P=\p{j}$ will be written simply \p{j}.)

\paragraph*{Manifested events.}
An \emph{event} is an occurrence at which at least one variable of the system takes on a
value randomly; this is brought about by a physical interaction of the system with its
exterior. \emph{Which} variable takes on a value randomly depends on the details of the
physical interaction, or \emph{manifestation}; at each event, then, a particular variable
is manifested. The dynamics of a probabilistic system deals with a (time) sequence of
events.%
\footnote{%
Both quantum mechanics and my card system are probability theories of sequences of
events; neither can be treated as a probability theory of values, because in neither can
the set of all value propositions be given a Boolean logical structure.
} %

\paragraph*{The v-state.}
An event is described by the values which have occurred; I will call this description
the value state (the ``\emph{v-state}''); the theory of the system yields probabilities
for the various branches, or v-states, possible for the event.

\paragraph*{Preparation and the p-state.}
A \emph{preparation} is an event process which erases any effect of the system's prior
history on the probabilities of succeeding events. A preparation of a system determines
the system's probability- or preparation-state (the ``\emph{p-state}''), a function which
allows the calculation of probabilities of every possible succeeding sequence of events.
The p-state is to be distinguished from the ``state'' of classical physics (to which, in
a sense, the v-state corresponds); it does not describe what is, but only the
probabilities of what might be. Because different preparations may result in the same
p-state, a p-state is implied by, but does not specify, a preparation. In quantum
mechanics, the p-state is equivalent with the statistical operator of the system. Thus
\begin{equation}
\text{Preparation $\Rightarrow$ p-state 
 $\stackrel{\text{QM}}{\Longleftrightarrow}$ statistical operator}.
\end{equation}
Though the p-state is determined by the preparation, it may then change according to a
deterministic dynamics (the \Schrodinger equation, for example); it does not, however,
change according to the outcomes of events (occurrences). We will write probability
expressions with the p-state (say $s$) as a subscript: \Pr[s]{\cdot}.

\paragraph*{Event sequence notation.}
An event's ordinal position in a sequence of events will be denoted by a superscript in
brackets: The event $E$ followed by the event $F$ is denoted $E^{[1]}\wedge F^{[2]}$.
Because this notation is rather awkward, we introduce the following simplifications
which will allow us to avoid the use of explicit ordinal superscripts for the most part:
\par\noindent(a) When the terms in the probability expressions are in the ``natural''
order and no ambiguity arises, the sequence ordinals will be dropped; thus \Pr[s]{\p{j}}
always means \Pr[\sz]{\p{j}^{[1]}}, and \Prob[s]{\q{k}}{\p{j}} always means
\Prob[\sz]{\q{k}^{[2]}}{\p{j}^{[1]}}. \par\noindent(b) In probability expressions
involving a conjunction such as \Pr[\sz]{x^{[1]}\wedge y^{[2]}}, we introduce the symbol
$\andthen$, ``and then,'' which implies the sequential order of conjunction:
\begin{equation}\label{E:AndThenDef}
 x\andthen y\DefEq x^{[1]}\wedge y^{[2]}.
\end{equation}
Then \Pr[s]{x\andthen y} always means \Pr[\sz]{x^{[1]}\wedge y^{[2]}}. \par\noindent(c)
All other orders of occurrence in probability expressions will require the explicit use
of the ordinal superscripts---\eg ``retrodiction,'' \Prob[\sz]{x^{[1]}}{y^{[2]}}.

\paragraph*{Filtered preparation.}
The conditional probability \Prob[s]{\cdot}{x} refers to the probability distribution of
the subset of the preparations which in the succeeding event satisfied the proposition
$X=x$; this distribution is the same as the distribution of a preparation consisting of
$s$ followed by the filter which passes only that subset $X=x$:
\begin{equation}\label{E:Filter}
  \Prob[s]{\cdot}{x}=\Pr[s\andthen x]{\cdot}.
\end{equation}

\paragraph*{``Observation.''}
Throughout this paper I will use the term ``observation'' for its simplicity and
familiarity. But it is too easy to infer from the use of this term the existence of an
observer, which connotes human conscious involvement and a concomitant collection of
metaphysical difficulties. As a physicist, not a metaphysicist, I always mean by
``observation'' only the minimal physical interactions necessary to assure the
occurrence of an event; presumably, a sufficiently clever human would then be able to
observe the value manifested in that event. Further, in probabilistic systems, that
which is ``observed'' is in most cases (think of flipping a coin, drawing a card,
passing a spin system through a Stern-Gerlach device) given its value by the very
process of ``observation'' (thus the scare-quotes, which, the point having been made, I
henceforth drop).

\section{Compatibility and interference}\label{S:Compatibility}

\subsection{Compatibility; \nojointP}\label{SS:Compatibility}
Here is a formal definition of compatibility (informally stated in \nojointP; recall the
notation introduced in \RefEqn{E:AndThenDef}):

\smallskip\noindent\textbf{Definition.~}The two variables $P$ and $Q$ are
\emph{compatible} \IFF
\begin{equation}\label{E:Compatibility}
 \Pr[s]{\p{j}\andthen\q{k}}=\Pr[s]{\q{k}\andthen\p{j}}
\end{equation}
for all indices, for every preparation state $s$.%
\footnote{%
Incompatibility does not arise in elementary probability texts, in which the usual
elementary examples of sequences are the drawing of balls from a urn or cards from a
deck, almost always done either \emph{with} replacement or \emph{without} replacement;
in either case the probability is independent of the order of occurrence. However, there
are many other replacement schemes (\eg ``replace if red, discard if green'') which do
not lead to this symmetry.}
\smallskip

The use of the quantum-mechanical term \emph{compatibility} for this classical definition
is not arbitrary: The following theorem establishes that this definition is
\emph{equivalent} with quantum mechanics' commuting-operators definition.

\par\smallskip\noindent\textbf{Theorem.~}Two variables $P$ and $Q$ of a quantum system
are compatible, $\Pr[s]{\p{j}\andthen\q{k}}$ $=\Pr[s]{\q{k}\andthen\p{j}}$ for all $j,
k$, and $s$ \IFF their corresponding operators \Op{P} and \Op{Q} satisfy
$\Op{P}\Op{Q}=\Op{Q}\Op{P}$.

\par\smallskip\noindent\emph{Proof:~} Expressing \RefEqn{E:Compatibility}, the
compatibility of the variables $P$ and $Q$, in quantal terms (utilizing the usual
``sandwich'' form for the probability of successive events, with the proposition $R=r_j$
being represented by the 1-projector $\Proj{r_j}\equiv\proj{r_j}$), we have
\begin{equation}\label{E:CompatibilityQ}
 \Trace{\bRho\,\Proj{\p{j}}\Proj{\q{k}}\Proj{\p{j}}}=%
 \Trace{\bRho\,\Proj{\q{k}}\Proj{\p{j}}\Proj{\q{k}}}.
\end{equation}
\RefEqn{E:Compatibility} holds for all p-states $s$, hence \RefEqn{E:CompatibilityQ}
must hold for all statistical operators $\bRho$, so
\begin{equation}\label{E:triple}
 \Proj{\p{j}}\,\Proj{\q{k}}\,\Proj{\p{j}}=\Proj{\q{k}}\,\Proj{\p{j}}\,\Proj{\q{k}}.
\end{equation}
Introduce\cite{Farina93} the operator
$\Op{C}\equiv\Proj{\p{j}}\,\Proj{\q{k}}-\Proj{\q{k}}\,\Proj{\p{j}}$. By
\RefEqn{E:triple}, $\OpAdj{C}\Op{C}$ is the zero operator, hence $||\Op{C}\ket{x}||^2=0$
for all \ket{x}, hence \Op{C} is the zero operator, \ie $\Proj{\p{j}}$ and $\Proj{\q{k}}$
commute; by \RefEqn{E:CompatibilityQ}, this implies the compatibility of $P$ and $Q$,
completing the implicative circle:  the compatibility of the variables $P$ and $Q$ is
equivalent with the commutativity of the basis projectors $\Proj{\p{j}}$ and
$\Proj{\q{k}}$.

The variables $P$ and $Q$ are represented by the hermitian operators \Op{P} and \Op{Q},
whose eigenexpansions are $\Op{P}=\Sum{t}\p{t}\Proj{\p{t}}$ and
$\Op{Q}=\Sum{t}\q{t}\Proj{\q{t}}$, respectively. As is well-known, the commutativity of
$\Proj{\p{j}}$ and $\Proj{\q{k}}$ is equivalent with the commutativity of \Op{P} and
\Op{Q}, which is thus equivalent with the compatibility of $P$ and $Q$.\QED

\subsection{Sharpness; \nosharpP}%
A variable is said to be \emph{sharp} in a given event if it has no statistical
dispersion; in probability terms, $\Pr[s]{\p{j}}\in\set{0,\,1}$. The condition
\ref{nosharpP} is expressed by
\begin{equation}\label{E:NoSharp}
 \Prob[s]{\q{k}}{\p{j}}\neq1
\end{equation}
(\ie if a system has been filtered to a sharp value of $P$, then a succeeding observation
of an incompatible variable $Q$ cannot yield a sharp value).

\subsection{Marginal probability; \margP}\label{SS:margprop}%
In the case of the sequences \set{\p{j}\andthen q}
 $\left(\text{\ie $\left\{{p_{j}}^{[1]}\wedge q^{[2]}\right\}$}\right)$,
the formula of marginal probability (\cf Appendix \ref{S:ElemProb}) would seem to imply
\begin{equation}\label{E:MargProbSeqBad}
 \Sum{t}\Pr[s]{\p{t}\andthen q}=\Pr[\sz]{q^{[2]}}.
\end{equation}
This expresses the erroneous assumption (based on the completeness of the values of a
variable) that $\Pr[s]{\big(\Orj{j}{\p{j}}\big)\andthen q}=\Pr[s]{\text{T}\andthen q}$ is
independent of the variable $P$ whose values are disjoined (or summed over)---that is,
that \Pr[\sz]{q^{[2]}} itself is defined. This is generally not the case (and is \margP
on the list of ``quantal'' phenomena). We will discuss this further in
\RefSec{S:Manifestation}, where we develop the correct form of the formula for marginal
probability of a sequence, \RefEqn{E:MargProbGood}.

This error, moreover, is often compounded by ignoring the sequential index, writing
\begin{equation}\label{E:MargProbFail}
  \Sum{t}\Pr[s]{\p{t}\andthen q}=\Pr[s]{q}.
\end{equation}
This seems to imply that $\Pr[\sz]{q^{[2]}}=\Pr[\sz]{q^{[1]}}$); this is generally
incorrect, even when \Pr[\sz]{q^{[2]}} is defined. Although the error is rather obvious,
it is exactly the error made by Margenau which led to his questioning the use of
classical probability in quantum mechanics; we discuss this further in
\RefSec{SS:Margenau}.

\subsection{Interference; \interfereP}%
In physical examples of classical wave systems (optics, acoustics, ocean waves, \dots),
the energy is additive for independent waves. As we analyze a wave process, we may
arbitrarily divide it into several alternate disjoint wave subprocesses, all having a
common endpoint. Interference is the difference (or, qualitatively, the existence of a
difference) between the value of the energy at the endpoint of the process and the sum
of the values of the energies of each of the several parallel subprocesses at that
endpoint. (If this difference is made a function of the endpoint, we refer to an
interference \emph{pattern}.) Interference is not itself a directly observable
phenomenon; rather, it is an artifact of the analysis of the physical system, and is
defined only in relation to the particular analytical decomposition. For example, in the
double-slit apparatus there is interference between the left and right slits, but there
is no interference between the upper halves of the two slits and their lower
halves---though the resulting pattern on the screen is the same in either case.

But classical wave interference is not directly applicable to quantum mechanics (nor to
any probabilistic system): theories of probabilistic systems predict the probabilities
of the occurrence of a value of a variable, not the value itself, so the energy can't be
used in the definition. Of course, there are quantum systems so similar to the classical
cases, both physically and mathematically (\eg the atomic Young apparatus) that the idea
of interference transferred rather directly, without formal definition; however, for many
other situations (\eg Wigner's recombining Stern-Gerlach apparatus), the analogy is much
less direct. Though it seems we all ``know it when we see it,'' still there is need for
an explicit definition of a generalization of interference to probabilistic systems, one
expressed in terms not restricted to the quantum formalism.

The probability of the disjunction of disjoint subprocesses is additive, thus the wave
concept of interference is generalized in a natural way to probabilistic phenomena by
giving the the role of the \emph{wave energy at the endpoint} to the \emph{probability
of the process}: In a probabilistic system, interference is the difference between the
probability of the process and the sum of the probabilities of the subprocesses. Based
on this line of thought, we offer the following definition of interference in
probabilistic systems:
\par\smallskip\noindent\textbf{Definition.~} Given an event \EvP compatible
with $P$ for which, for all preparation states $s$,
\begin{subequations}\label{E:IntDef}
\begin{equation}\label{E:IntCrit1}
\Pr[s]{\EvP}=\sum_{t\in D}\,\Pr[s]{\p{t}},
\end{equation}
the \emph{interference} of \EvP with respect to $\{\p{j}\,|\,j\in D\}$ is
\begin{equation}\label{E:IntCrit2}
I(\EvP,\,\{\p{j}\,|\,j\in D\},\,s,\,q)=%
 \Pr[s]{\EvP\andthen q}-\sum_{t\in D}\,\Pr[s]{\p{t}\andthen q}.
\end{equation}\end{subequations}

Thus (in the case $D=\set{1,\,2}$) \EvP \emph{appears} to be the disjunction
$\p{1}\vee\p{2}$---but this ``disjunction'' doesn't distribute: $\EvP\wedge
q\not\equiv(\p{1}\wedge q)\vee(\p{2}\wedge q)$. This phenomenon of quantum interference
(\interfereP on the list of ``quantal'' properties) was described by
Feynman\cite{FeynmanVolIII} as the ``heart'' of quantum mechanics, its ``only mystery.''%
\footnote{Feynman's statement was made prior to Bell's publications, hence the
singularity of the mystery; but nothing about Bell's insights makes interference any
\emph{less} mysterious.} %
In quantum mechanics, interference arises exactly in the case that \EvP appears to be
the disjunction $\p{\,1}\Or\p{\,2}$, but the apparent alternatives \p{\,1} and \p{\,2}
are physically indistinguishable (as in, for example, the atomic double-slit apparatus)
and the apparent disjunction fails to distribute.%
\footnote{%
This failure of distribution of disjunction is, of course, one basis for the introduction
of ``quantum logic.'' Another (related) reason is described in footnote \ref{FN:patch}.
} %

\section{A classical system exhibiting the properties \theseP}\label{S:Mimick}
Here is an example of an entirely classical probabilistic system which illustrates the
ordinary nature of much of ``quantum probability'': incompatibility, the non-existence of
dispersion-free ensembles, the ``failure'' of the marginal-probability formula, and
interference.

In order to make this system dramatically non-quantal, I construct it using playing
cards. About these cards the reader need only know that each carries two marks, the
``face'' and the ``suit'' (traditionally with names such as King, Queen, \dots, and
Spades, Hearts, \dots, respectively), and that the suits are marked in two colors, red
(Hearts and Diamonds) and black (Spades and Clubs). I treat face, suit, and color as
system variables; each variable is ``observed'' (given a value) according to the rules
of observation described below.

Interference may arise from a manifestation process which treats several values of a
variable identically, as in the case of a degenerate value in quantum mechanics. In this
card example, a natural choice for degeneracy is the \emph{color} of the suit. The color
of a card may be observed in either of the following ways: We may observe the suit and
report the color (the failure to have a value for the suit is a matter of
ignorance---which is what we mean by ``we ignored the suit''); in such a case,
\RefEqn{E:IntCrit2} vanishes. Alternatively, we may observe the color using a
manifestation which processes the red suits identically, but differently from the black
suits; in our example this leads to interference (the nonvanishing of
\RefEqn{E:IntCrit2}).

\par\smallskip\noindent\textbf{System \SysS}\par
\noindent A deck of playing cards, having two variables, \Face and \Suit, each of which
may take on a disjoint set of values: \K, \Q, \J,  and \S, \H, \D, respectively. The
variable \Color (a function of \Suit) takes on the values \R (red) and \B (black).
Duplicate cards are allowed, with the restriction that each \Face and each \Suit
appear in equal numbers (so their \emph{a priori} probabilities are equal).%
\footnote{For example, in the case of two-valued variables we might use the deck
\set{\K\S,\K\S,\K\H,\Q\S,\Q\H,\Q\H}: three of each value.} %
The variable under consideration (\Face, \Suit, or \Color) is denoted by $P$; its values
are \set{\p{j}}.
\par\smallskip\textbf{Observation:} To observe a variable $P$:\smallskip
\par\qquad\;1. Shuffle the subdeck.
\par\qquad\;2. Report \p{j}, the $P$-value of the subdeck's top card.
\par\qquad\;3. Construct a new subdeck consisting of all cards for which $P=\p{j}$.
\begin{quote}
\Eg to observe \Face, shuffle the subdeck and report the \Face-value of its top card,
\Q, say, then construct a new subdeck consisting of all the \Q's in the deck.

To observe \Color, shuffle the subdeck; if, say, the top card's \Suit-value is \H (a red
card), report the \Color \R, then construct a new subdeck consisting of all the deck's \R
cards: all its \H's and all its \D's.
\end{quote}

\par\smallskip\textbf{Preparation:} To prepare the system in the state ``the value of
$P$ is \p{j}'':\smallskip
\par\qquad\textbf{repeat}\
\par\qquad\quad1. Observe any other, incompatible, variable $Q$ (ignoring the result).
\par\qquad\quad2. Observe $P$.
\par\qquad\textbf{until} $P=\p{j}$
\begin{quote}
(This results in a subdeck consisting of all cards for which $P=\p{j}$.)
\end{quote}

The system \SysS exemplifies incompatibility (\ref{incompP}): The processes for observing
\Suit and \Face cannot be carried out simultaneously. For example, if the card on the
top of the subdeck is $\Q\H$, then the reporting of \Face would require the construction
of a new subdeck consisting of all the \Q's, while the reporting of \Suit would require
the construction of a new subdeck consisting of all the \H's. It is impossible to carry
out these two constructions simultaneously (unless the deck contains no \H's and no \Q's
other than the $\Q\H$).%
\footnote{Of course, one might cheat and look at both values as marked on the card;
however, it is impossible to follow both subdeck-construction rules, and thus the cheat
would be meaningless, irrelevant to the behavior of the system---as irrelevant as, say,
having just determined the $z$-component of spin, flipping a coin to ``determine'' the
$x$-component.}%

The statistical behavior of \SysS is summarized in \RefEqn{E:results}, in which I use
the following notation: $P$, $Q$, $X$, and $Y$ are system variables; $P$ and $Q$ are
different variables, with possible values \set{\p{j}} and \set{\q{k}}, respectively,
while $x$ and $y$ are (not necessarily distinct) values of the (not necessarily
different) variables $X$ and $Y$.  (In the example, the variables are \Face or \Suit,
their values are \K, \Q,\dots, \S, \H,\dots.) The equalities in \RefEqn{E:results} hold
for all value indices; the inequalities hold for at least some values. The system is
prepared in the state $s$.
\begin{subequations}\label{E:results}\allowdisplaybreaks\begin{alignat}{2}
   &\Prob[\sz]{\p{k}^{[2]}}{\p{j}^{[1]}}=\KDelta{j}{k}%
   &\quad&\text{(repeatability);}\label{E:repeatable}\\
   &\Prob[s]{\q{k}}{\p{j}}=\Prob[s]{\p{j}}{\q{k}}%
   &&\text{(reciprocity);}\label{E:reciprocity}\\
   &\Pr[s]{x\andthen y}=\Pr[x]{y}\Pr[s]{x}%
   &&\text{(Markovian);}\label{E:Markov}\\
   &\Pr[s]{\p{k}\andthen\p{l}}=\Pr[s]{\p{l}\andthen\p{k}}%
   &&\text{(self-compatibility);}\label{E:compatible}\\
   &\Pr[s]{\p{k}\andthen\q{l}}\neq\Pr[s]{\q{l}\wedge\p{k}}%
   &&\text{(incompatibility);}\!\!\label{E:incompatible}\\
   &\Prob[s]{\q{k}}{\p{j}}\neq1
   &&\text{(if P is sharp, Q is not);}\label{E:notSharp}\\
   &(\exists\,j\suchthat\,\Pr[s]{\p{j}}\notin\set{0,1})\Longrightarrow
   &&\text{(marginal probability\qquad\quad}\notag\\
   &\qquad\qquad\textstyle{\sum_{t}}\Pr[s]{\p{t}\andthen\q{k}}\neq\Pr[s]{\q{k}}%
   &&\text{\quad\quad formula ``fails'');}\label{E:margFails}\\
   &\text{($\alpha$)}\;\Pr[s]{\R}=\Pr[s]{\H\vee\D}\;\forall\,s,\text{ but}%
   &&\text{(the interference of $\R$\qquad\qquad}\notag\\
   &\text{($\beta$)}\;\Pr[s]{\R\andthen\q{k}}\neq\Pr[s]{\big(\H\vee\D\big)\andthen\q{k}}%
   &&\text{\quad\quad relative to $\H\vee\D$).}\label{E:interference}
\end{alignat}
\end{subequations}
(These results are derived in Appendix \ref{S:development} for any number of variables
with any number of values.)

From \RefEqn{E:repeatable} we see that, following the observation of, say, \Face, the
ensemble is sharp in \Face: the observation is \emph{repeatable}. In quantum mechanics,
\RefEqns{E:reciprocity} and \eqref{E:Markov} hold for nondegenerate values.
(\RefEqn{E:Markov} may be written equivalently as
\begin{equation}
 \Prob[s]{y}{x}=%
 \begin{cases}
   \Pr[x]{y},
         &\text{if }\Pr[s]{x}\neq0\tag{\ref{E:Markov}$^{\prime}$}\\
   \text{undefined}&\text{otherwise;}
  \end{cases}
\end{equation}
filtering a manifestation to a specific result erases any ``memory'' of the earlier
preparation state.)

\RefEqn{E:compatible} shows that, as in quantum mechanics, the variables of this system
are compatible with themselves. \RefEqns{E:incompatible}--\eqref{E:interference} are the
archetypal ``quantal'' effects \statP.

\RefTables{Tab:Prob}--\ref{Tab:Interference} illustrate results for a simple version of
the system, involving just two variables, each with three values.  \RefTable{Tab:Marg}
shows an apparent failure of completeness, and \RefTable{Tab:Interference} shows an
apparent failure of the distributive rule; we discuss these ``failures'' in the
succeeding section.

\begin{table}[t]%
\begin{tabular*}{\columnwidth}{@{\extracolsep{\fill}}cccccc}\hline\hline
   \strt&&\S&\H&\D&\\\hline
   \strt&\K&0.1&0.4&0.5&\\
   \strt&\Q&0.4&0.5&0.1&\\
   \strt&\J&0.5&0.1&0.4&\\\hline
\end{tabular*}
\begin{tabular*}{\columnwidth}{@{\extracolsep{\fill}}ccc}
 \strt & \Prob{\Suit_k^{[1]}}{\Face_j^{[0]}}= \Prob{\Face_j^{[1]}}{\Suit_k^{[0]}}&\\%
 \hline\hline
\end{tabular*}
\caption{\protect The basic probabilities; note that there are no dispersion-free states
(\nosharpP). (Multiplying each table entry by (a multiple of) 10 gives the number of
duplicates of that card.) } \label{Tab:Prob}
\end{table}

The system \SysS exhibits the phenomena \theseP as a result of the laws of probability
and an ``intelligent'' choice of selection rules and cards---it does not simulate them
(in the sense that a system constructed using Newtonian mechanics and an ``intelligent''
choice of parameters and initial conditions \emph{exhibits} the orbital phenomena of the
solar system; an orrery, or a combination of observational data and Kepler's Laws,
\emph{simulates} those phenomena). Further, \SysS does not approximate the specific
quantitative results of quantum mechanics (which, of course, also satisfies \statP);
\SysS is neither a model nor a mechanization of quantum mechanics.

\section{Manifestation}\label{S:Manifestation}%
\subsection{Apparent problems with completeness}
A naive consideration of the completeness of the values of a variable would suggest
that, since \set{\K,\Q,\J} exhausts the possibilities of \Face (in the decks being
considered in \SysS), it must be that $\K\vee\Q\vee\J=\text{T}$. Similarly, it must be
that $\S\vee\H\vee\D=\text{T}$. And this is correct in many ways; for example,
$\Pr[x]{\K\vee\Q\vee\J}=\Pr[x]{\K}+\Pr[x]{\Q}+\Pr[x]{\J}=1$ for any preparation $x$.
However, note that
\begin{subequations}\label{E:abcK}\begin{align}
 &\Pr[{\Q}]{\big(\K\vee\Q\vee\J\big)\andthen\K}=0,\label{E:KQJK}
 \intertext{but}
 &\Pr[{\Q}]{\big(\S\vee\H\vee\D\big)\andthen\K}\notag\\%
 &\qquad=\Pr[{\Q}]{\S\andthen\K}+\Pr[{\Q}]{\H\andthen\K}+\Pr[{\Q}]{\D\andthen\K}\notag\\
 &\qquad=\Pr[\S]{\K}\Pr[\Q]{\S}+\Pr[\H]{\K}\Pr[\Q]{\H}+\Pr[\D]{\K}\Pr[\Q]{\D}\geq0
\label{E:SHDK}
\end{align}\end{subequations}
(using \RefEqns{E:repeatable}--\eqref{E:Markov}). This can vanish only in a specially
selected deck such that the \K's and the \Q's share no suit---not the general case. But
these expressions would necessarily be equal were both ${\K\vee\Q\vee\J}$ and
$\S\vee\H\vee\D$ equal to $\text{T}$.

Thus $\K\vee\Q\vee\J$ and $\S\vee\H\vee\D$ cannot both be true (hence equal) at the same
time. But if $\K\vee\Q\vee\J\neq\text{T}$ it must be that \Face has no value at all. This
is exactly the case in our cardgame example: The observations of \Face and \Suit are
incompatible processes, hence \Face and \Suit cannot have values
simultaneously.%
\footnote{\label{FN:patch}The erroneous identification of the two probability-1
disjunctions $\Orj{j}{\p{j}}$ and $\Orj{k}{\q{k}}$ (of incompatible variables $P$ and $Q$
respectively) has mislead us to quantum logic by suggesting that we ``patch together''
the Boolean proposition lattices of $P$ and $Q$ into a non-modular lattice, by
identifying their least elements and identifying their greatest elements.} %

\subsection{Manifestation}
But if a variable does not always have a value, there must be a process which causes a
variable $P$ to take on one of its values; I call that process the \emph{manifestation}
of $P$ and denote it $M_P$. Each event in this probability system of variable values must
be a manifestation. The probability $\Pr[x]{\S\andthen\K}$, for example, has meaning
only in the context of the \emph{manifestation history} $M_{\Suit}\andthen M_{\Face}$ as
the condition of the probability: $\Prob[x]{\S\andthen\K}{M_{\Suit}\andthen M_{\Face}}$.

\begin{table}[t]%
\begin{tabular*}{\columnwidth}{@{\extracolsep{\fill}}cccccc}\hline\hline
   \strt&&\S&\H&\D&\\\hline
   \strt&\K&~0.09&~0.36&~0.45&\\
   \strt&\Q&-0.16&-0.20&-0.04&\\
   \strt&\J&-0.25&-0.05&-0.20&\\\hline
\end{tabular*}
\begin{tabular*}{\columnwidth}{@{\extracolsep{\fill}}ccc}
 \strt &$\Prob{\Face_k^{[1]}\wedge\Suit_l^{[2]}}{\K}-%
                    \Prob{\Suit_l^{[1]}\wedge\Face_k^{[2]}}{\K}$&\\%
 \hline\hline
\end{tabular*}
\caption{\protect\strt  Incompatibility (\nojointP). Note that the two variables are
incompatible in every value, and that the incompatibilities are not symmetric.}
\label{Tab:Incompat}
\end{table}

With the understanding that this manifestation history is a necessary part of the
probability expression, it is often safe to leave it implicit---but not always: If we
make the manifestations in \RefEqn{E:abcK} explicit, we may safely identify the
disjunctions as true; for example, we may write \RefEqn{E:SHDK} as
\begin{equation}\label{E:K}
 \Prob[{\Q}]{\big(\S\vee\H\vee\D\big)\andthen\K}{M_{\Suit}\andthen M_{\Face}}%
 =\Prob[{\Q}]{\text{T}\andthen\K}{M_{\Suit}\andthen M_{\Face}}.
\end{equation}
The $M_{\Face}^{[2]}$ may be made implicit (``dropped'') without harm, but
\Pr[{\Q^{[0]}}]{\K^{[2]}} would be ambiguous regarding the occurrence at $[1]$, so the
$M_{\Suit}^{[1]}$ (and $M_{\Face}^{[1]}$, in \RefEqn{E:KQJK}) must be left explicit.
Thus, we may write \RefEqns{E:KQJK} and \eqref{E:SHDK} as
\begin{subequations}\label{E:KgQM}\begin{align}
 &\Prob[{\Q}]{\K}{M_{\Face}}=0\\
 &\Prob[{\Q}]{\K}{M_{\Suit}}\geq0;
\end{align}\end{subequations}
that these may differ is no surprise.

\subsection{Congruence with the manifestation history}
Let us generalize this. Accounting for the manifestation history and the disjointness of
\set{\p{j}},
\begin{equation}
\Sum{t}\Pr[s]{\p{t}\andthen q}=%
  \Prob[s]{\big(\Orj{t}{\p{t}}\big)\andthen q}{M_P\andthen M_Q}.
\end{equation}
Conditionalize the right side probability on the disjunction \Orj{t}{\p{t}} (whose
probability is 1), and note that, while $M_Q^{[2]}$ may be made implicit without
ambiguity, the manifestation $M_P^{[1]}$ must be denoted explicitly, thereby obtaining
the correct form of the marginal-probability formula for sequences involving incompatible
variables:
\begin{equation}\label{E:MargProbGood}
   \Sum{t}\Pr[s]{\p{t}\andthen q}=\Prob[s]{q}{M_P}.
\end{equation}
The right side expresses \margP explicitly: \emph{Though we have ignored the value of an
observed variable, we may not ignore the fact of that variable's observation.}

This discussion leads to the rule
\begin{quote}
\emph{The probability of an event-sequence necessarily includes the manifestation
history in the condition of the probability expression (and the event-sequence must be
congruent with that manifestation history)}.%
\footnote{The event-sequences congruent with a given manifestation history form a Boolean
event space.}
\end{quote}

The simultaneous manifestation $M_{\Face}^{[1]}\wedge M_{\Suit}^{[1]}$ is impossible (as
already pointed out, in \RefSec{S:Mimick}). An impossible condition leads to an undefined
conditional probability (\cf \RefEqn{E:CondProb}); thus
$\Pr[x]{\K\wedge\S}\equiv\Prob[x]{\K^{[1]}\wedge\S^{[1]}}{M_{\Face}^{[1]}\wedge
M_{\Suit}^{[1]}}$ is undefined---disallowed logically, not by decree. Generally,
\begin{quote}
\emph{The simultaneous manifestation of the values of several incompatible variables is
impossible, hence the probability of the simultaneous conjunction (or disjunction) of
their values is meaningless---their joint probability distribution does not exist.}
\end{quote}

\begin{table}[t]%
\begin{tabular*}{\columnwidth}{@{\extracolsep{\fill}}cccccc}\hline\hline
   \strt&&$\K^{[1,2]}$&$\Q^{[1,2]}$&$\J^{[1,2]}$&\\\hline
   \strt&$\K^{[0]}$&-0.90&~0.40&~0.50&\\
   \strt&$\Q^{[0]}$&~0.40&-0.50&~0.10&\\
   \strt&$\J^{[0]}$&~0.50&~0.10&-0.60&\\\hline
\end{tabular*}
\begin{tabular*}{\columnwidth}{@{\extracolsep{\fill}}ccc}
  \strt 
    &$\Prob{\big(\S\vee\H\vee\D\big)^{[1]}\wedge\Face_k^{[2]}}{\Face_j^{[0]}}\,-\,%
      \Prob{\Face_k^{[1]}}{\Face_j^{[0]}}$&\\%
  \hline\hline
 \end{tabular*}
\caption{\protect\strt The effect of ignoring a prior observation of \Suit---an apparent
failure of the marginal-probability formula (\margP). } \label{Tab:Marg}
\end{table}

\subsection{Significance of manifestation}
In a probabilistic system, events are physical processes, interactions with the exterior
of the system; the nature of the event, in particular the identity of the variable which
is randomly affected (``takes on a value'') in an event, depends on the details of the
physical interaction. We have expressed this as \emph{manifestation}, a necessary
component of any discussion of processes in physically realistic probabilistic systems
(quantal or other). Manifestation is (due to) the physical interaction of the system
under consideration with its exterior; in the absence of such interaction, there is no
event, no probabilistic branching. Manifestation implements Bohr's dictum that the entire
situation must be taken into account; while Bohr's requirement is (merely) metaphysical,
the requirement that manifestation be taken into account is a logical consequence of this
analysis.

Both quantum mechanics and the theory describing \SysS are probability theories of
\emph{sequences of events}, not of the events themselves. These sequences' probabilities
are defined in terms of a classical Boolean probability space whose elements are the
sequences congruent with a given manifestation history. When we carry out all analysis in
terms of the sequence-element, no anomalies arise. It may appear that these
probabilities fail the Kolmogorov postulates if we forget their contextuality: that
congruence. Taking into account the physical situation---the manifestation
history---properly restricts the choice of the sequence of events, guaranteeing a Boolean
probability space. As we see in these ordinary systems, this contextuality is a
respectable, non-subjective property of a probabilistic system with more than one
variable.

It is common in probability applications to ignore (``sum out'') the outcome of an
event; we've seen, however, that in general \emph{we may not ignore the fact that an
outcome was ignored}---we must take into account the fact of the manifestation of the
ignored variable to avoid numerous anomalies, such as the apparent failure of the
formula for marginal probability, the appearance of non-Boolean probability structures,
and difficulties such as the ``Curious'' results of \RefCite{AlbertAD85} (which I discuss
in \RefCite{Kirkpatrick:ThreeBox}).

\subsection{Margenau, marginal probabilities, and manifestation}\label{SS:Margenau}%
For a quantum mechanical system prepared in the state $\Psi$, we have
\begin{subequations}\begin{equation}\label{E:M1a}
 \Pr[\Psi]{\q{k}}=\abs{\braket{\q{k}}{\Psi}}^2.
\end{equation}
But also, for the non-degenerate values \set{\p{j}} (whose occurrence, in quantum
mechanics, is Markovian: $\Prob[\Psi]{\q{k}}{\p{j}}=\Pr[\p{j}]{\q{k}}$), we have
\begin{equation}\label{E:M1b}
 \Sum{t}\Pr[\Psi]{\p{t}\andthen\q{k}}%
  =\Sum{t}\Prob[\Psi]{\q{k}}{\p{t}}\Pr[\Psi]{\p{t}}%
  =\Sum{t}\abs{\braket{\q{k}}{\p{t}}}^2\abs{\braket{\p{t}}{\Psi}}^2.
\end{equation}\end{subequations}

According to the conventionally accepted formula for marginal probability
(\RefEqn{E:MargProb}), these should be equal; however, as Margenau\cite{Margenau63a}
pointed out, they are not:
\begin{equation}
 \Sum{t}\abs{\braket{\q{k}}{\p{t}}}^2\abs{\braket{\p{t}}{\Psi}}^2\neq%
        \abs{\braket{\q{k}}{\Psi}}^2.
\end{equation}
Margenau interpreted this as establishing the failure of classical probability within
quantum mechanics.

\begin{table}[t]%
\begin{tabular*}{\columnwidth}{@{\extracolsep{\fill}}cccccc}\hline\hline
   \strt&&$\K^{[2]}$&$\Q^{[2]}$&$\J^{[2]}$&\\\hline
   \strt&$\K^{[0]}$&-0.005&~0.020&-0.015&\\
   \strt&$\Q^{[0]}$&~0.020&-0.080&~0.060&\\
   \strt&$\J^{[0]}$&-0.015&~0.060&-0.045&\\\hline
\end{tabular*}
\begin{tabular*}{\columnwidth}{@{\extracolsep{\fill}}ccc}
 \strt &$\Prob{\R^{[1]}\wedge\Face_k^{[2]}}{\Face_j^{[0]}}-%
        \Prob{(\H\vee\D)^{[1]}\wedge\Face_k^{[2]}}{\Face_j^{[0]}}$&\\%
  \hline\hline
 \end{tabular*}
\caption{\protect\strt  Numerical results demonstrating interference (\interfereP)---the
apparent failure of the distributive rule
 $(p\vee q)\wedge r=(p\wedge r)\vee(q\wedge r)$. } \label{Tab:Interference}
\end{table}

However, as exemplified in \RefSec{S:Mimick}, this ``failure'' occurs in ordinary
probability settings. The marginal-probability formula for sequences of events is
correctly given by \RefEqn{E:MargProbGood}; \RefEqn{E:M1b} is to be equated, not with
\RefEqn{E:M1a}, but with \Prob[\Psi]{\q{k}}{M_P}. Now (\cf \RefEqn{E:Filter})
\begin{equation}
  \Prob[\Psi]{\q{k}}{M_P}=\Pr[\Psi\andthen M_P]{\q{k}}=\Trace{\bRho_P(\Psi)\,\proj{\q{k}}},
\end{equation}
with $\bRho_P(\Psi)$, the p-state after preparation in $\Psi$ followed by manifestation
of $P$, given by
\begin{equation}
  \bRho_P(\Psi)=\Sum{t}\proj{\p{t}}\,\ket{\Psi}\bra{\Psi}\,\proj{\p{t}};
\end{equation}
thus
\begin{equation}
  \Prob[\Psi]{\q{k}}{M_P}=%
   \Sum{t}\abs{\braket{\q{k}}{\p{t}}}^2\abs{\braket{\p{t}}{\Psi}}^2,
\end{equation}
in accordance with the correct marginal-probability formula \RefEqn{E:MargProbGood}. No
failure of classical probability within quantum mechanics arises here---only a failure
to apply classical probability correctly.

\section{Indeterminate values}\label{S:Indeterminate}%
There is a further property which has played a major role in the interpretation of
quantum mechanics---that of the ``nonreality'' of quantum systems, the
value-indeterminate nature of variables:
\begin{enumerate}
\renewcommand{\theenumi}{Q\arabic{enumi}}\setcounter{enumi}{\value{holdenumi}}
\item The variables are value-indeterminate, having no value except as one arises upon
observation.\label{indetP}
\end{enumerate}
Value-indeterminacy is suggested by \BohrP---Bohr included it in his metaphysical
principle of \emph{complementarity}. In quantum mechanics, it is suggested even more
strongly by results arising directly out of the formalism.\cite{Bell66,KochenSpecker67}
Although quite disturbing to classically-trained physicists with a bias toward
determinism (that is, all of us), it is not in fact particularly strange---merely by
making the system \SysS of \RefSec{S:Mimick} truly nondeterministic, we obtain a real
system (\SysSvi, below) which is seen by internal analysis to be nonrealistic
(value-indeterminate).

First, some definitions:
\par\smallskip\noindent\textbf{Definition~}\emph{Value-Determinate System}. At every
instant there exists, for each variable of the system, a value which would be the result
were that variable to be the next observed.\smallskip

\noindent(This does not require that, at such instant, the variable physically have the
value that it will, if observed, display.)

\par\smallskip\noindent\textbf{Definition~}\emph{Deterministic System}. The outcome of a
future event is a function of the values of the variables of the system at an earlier
time. (This is \emph{future}-determinism, all we need for the following development.)
\par\smallskip
In a deterministic system the outcome of the next observation of a variable is determined
by the present state, so the result of that future observation has a value now: \emph{A
deterministic system is necessarily value-determinate.}

\subsection{Examples of value-determinacy and -indeterminacy}\label{SS:exInd}
In the system \SysS, introduced and discussed in \RefSec{S:Mimick}, the shuffling of the
deck was taken to be deterministic (\eg mechanical); in that case, the outcome of the
next pick exists now---the variables are value-determinate. However, this determinism is
not necessary---the choice of the next card may be made nondeterministically, and this
may lead to value-indeterminacy. To implement this nondeterministic choice we may use
any of a number of physical systems with random behavior---nuclear decay, Josephson
junction tunneling, photons impinging on a beam splitter---to generate truly
nondeterministic random numbers to pick the cards. Thus, consider \SysSvi:

\par\smallskip\noindent\textbf{System \SysSvi}\par \noindent The system \SysSvi is
exactly as \SysS except that the deck is shuffled nondeterministically.\smallskip

The statistics of \SysSvi and of \SysS are identical, satisfying all of \RefEqn{E:results}.
\SysSvi is perfectly real (and easily constructed),%
\footnote{\SysSvi is not ``classical''; classical physics is the study of a strictly
deterministic world. Neither, however, does the presence of a Josephson junction, say,
make \SysSvi ``quantal''; no aspect of the theory of quantum mechanics need be used in
its analysis.} %
but it is not ``realistic''---it is clearly a value-\emph{in}determinate system: if the
most recently observed variable were, say, \Face, then whatever value would appear if
\Suit were to be observed in a moment \emph{does not exist} before that process of
observation, and will only be brought into existence by the nondeterministic shuffling of
the deck at the time of observation.

The value-indeterminacy of \SysSvi arises strictly from its nondeterministic card choice;
the systems \SysS and \SysSvi are otherwise identical. However, nondeterminism does not,
in and of itself, imply value-indeterminacy, as we see in \SysSvd:

\smallskip\noindent\textbf{System \SysSvd}\par\noindent The system \SysSvd is a
modification of \SysS, in which the manifestation rule is carried out in the order 2, 3,
1 (\ie shuffle last).
\begin{itemize}
\item[] \Eg observing \Face following the prior observation of $\Suit=\H$, the subdeck
consists of all the \H's. Report the \Face-value of its top card, \Q, say; create a new
subdeck consisting of all the \Q's, then shuffle it.
\end{itemize}
\SysSvd is value-determinate (whether the shuffling is deterministic or
nondeterministic): both \Face and \Suit have values (those of the top card) \emph{prior}
to an observation of either---although only one can be observed, the other being
disturbed by that observation. (The statistical properties of \SysSvd, including
interference, are identical with \SysS and \SysSvi.)

\subsection{Discussion: Value indeterminacy in quantum mechanics}
These examples show us that ``nonrealism'' is a straightforward possibility in
nondeterministic systems. However, in contrast with \statP, which are statistical, that
is, phenomenological, value-indeterminacy is an ontic property with no characteristic
empirical consequence. It can be demonstrated only by analysis of the workings of the
system: \SysS and \SysSvi have identical observable behavior, but one is
value-determinate, the other value-indeterminate. Furthermore, while nondeterminism is
necessary for value indeterminacy, as we have seen from the example of \SysSvd it is not
sufficient.

We should note here that interference depends neither on value-indeterminacy nor on
nondeterminism: it occurs in the value-determinate \SysSvd as well as in the
value-indeterminate \SysSvi, both nondeterministic, as well as in the deterministic and
value-determinate \SysS.

The problem of indeterminate values---the lack of ``reality'' of the values of
variables---has been a central difficulty for the interpretation of quantum mechanics:
\RefCite{Redhead89}, for example, concludes with ``In this book we have been mainly
concerned with the difficulties encountered by a simple-minded realism of possessed
values''; because of indeterminacy, \RefCite{dEspagnat95} considers reality to be
``veiled''; the consistent-histories interpretation and the various modal interpretations
all have as their central purpose the avoidance of value-indeterminate variables, while
Copenhagenism goes to the other extreme, metaphysically demanding value-indeterminacy
under the philosophical principle of complementarity.

However, in the ordinary, non-quantal system \SysSvi, the values are indeterminate---the
mechanism of this system makes it clear that a newly manifested variable \emph{had no
value} prior to its manifestation. Value-indeterminacy is a normal possibility in a
nondeterministic system (at a certain point in the manifestation process, a
nondeterministic choice brings one variable's value into existence, and, at the same
moment, pushes the other variable's value out of existence).%
\footnote{The classic example of indeterminate value is due to Aristotle: $B=$~``There
will be a sea-battle tomorrow.'' Then $B\vee\Not B$ is true, but, assuming
nondeterminism in human affairs, neither $B$ nor $\Not B$ has a truth value today.} %
Thus, value indeterminacy has no ``explanation'' beyond the the ontic fact of
nondeterminism: though we have complete understanding of their internal structure and
behavior, we gain no explanation of the nonrealism of our classical examples beyond the
analysis which demonstrates it.

We may caricature as the view of Heisenberg and of Bohr, respectively, that the
incompatibility of variables is \emph{epistemic} (it is merely that we have no
technique, even, perhaps, in principle, to observe at one instant the values of
incompatible variables) or \emph{ontic} (it is that they have no such values). Each of
these alternatives is, in fact, quite possible, as \SysS and \SysSvi illustrate.
Epistemic and ontic incompatibility are not other than the value-determinacy or
-indeterminacy of the incompatible variables.

The mystery of the indeterminate values of quantum mechanics is not to be resolved
through (unattainable) detailed knowledge of the system, nor by deep metaphysics; it is
not other than the mystery of nondeterminism.

\section{Conclusion}\label{S:Conclusion}%
Each of the phenomena \allP has been considered, by one author or another, to be
characteristic of quantum mechanics, inexplicable and unacceptable from the viewpoint of
classical physics. The problematic nature of these apparently quantal
properties---incompatibility, the non-existence of dispersion-free pure states,
interference, value-indeterminacy---seems to call for the ``interpretation'' of quantum
mechanics---but, unfortunately, not in any single interpretational direction: The
problems raised by \BohrP encouraged Bohr's complementarity principle and
``Copenhagenism''; the failure of the distributive rule of logic (\margP and
\interfereP) has given rise to various quantum probabilities and logics. The various
modal and consistent-histories interpretations arose primarily in order to solve the
``problem'' of quantal ``nonrealism,'' \indetP.%
\footnote{This oversimplification of the development of these interpretive systems is
not, I think, misleading in the present context.} %

There are, of course, interpretive difficulties with quantum mechanics other than those
associated with \allP: The issues of ``collapse'' and the Measurement Problem (alias
``\Schrodinger's Cat''), which have generated the decoherence approach; and the
impossibility, established by Bell, of reducing quantum mechanics (specifically, its
distant correlations) to classical-mechanical kinetic theory (which includes, of course,
any scheme based on systems of the sort presented in this paper). I have not
dealt with any of these issues in this paper.%
\footnote{However, I might comment that the first of these is greatly clarified by
recalling that a the state concept in a probabilistic system is categorically different
from that of a deterministic system; the second and third are similarly greatly
clarified by the recognition that a probabilistic system, which quantum mechanics
certainly is, may well be (though there is no phenomenological test) irreducibly
probabilistic, that is, nondeterministic, hence its variables may well be
value-indeterminate (\indetP)---in which case the usual derivations of the Bell conditions
fail, and the positivity requirement in Fine's derivation\cite{Fine82b} loses cogency.} %

But the phenomena \statP are, from a probability viewpoint, quite unexceptional, even
expected, and certainly comprehensible: they have all been exemplified in an ordinary
(``classical'') probabilistic system \SysS. The appearance of \statP in this
emphatically non-quantal system establishes decisively that none of these phenomena are
quantal.

Both quantum mechanics and \SysS (absolutely distinct from quantum mechanics) are
examples of probabilistic systems having more than one variable; because
\nojointP--\interfereP appear in both, we must conclude that these phenomena are
characteristic of (some subset of) probabilistic systems of several
variables, but are characteristic neither of \SysS nor of quantum mechanics.%
\footnote{%
Socrates' mortality is characteristic, not of Socrates, but of a subset of all beings,
the \emph{mortal} beings, of which he is a member. Although a study of the details of
Socrates' mortality may bear fruit (or hemlock), questions regarding the meaning of his
mortality---its ``interpretation''---must be directed at, and studied within the context
of, \emph{mortal beings}. } %
Nor are these phenomena in any way \emph{quantal} weirdness: any metaphysical problems
these phenomena may present are problems for the entirety of their subset of theories,
and require no special interpretations of quantum mechanics for the understanding of
their implications (although their appearance in this card game example rather removes
the sense there might be ``implications'' needing ``understanding'').

No explanation of the appearance of \statP in quantum mechanics is necessary beyond
noting that quantum mechanics is a probabilistic system which has more than one
variable. It may be that analysis of the workings of a system exhibiting  \statP will
shed light on the particulars of the mechanism by which they are expressed (though, in
quantum mechanics, seventy-five years of trying have yielded no such prize). That, in
fact, not much such light can or need be shed is supported by the theorem of
\RefSec{SS:Compatibility}: Given the Hilbert-space formalism of quantum mechanics,
statistical incompatibility of variables requires neither more nor less than the
noncommutativity of the variables' operators: there is no room for further explanation
(in the sense that, as mechanical conservation laws are sufficient to explain the
center-of-mass motion of colliding billiard balls, no room is left for further
explanation of that center-of-mass motion though study of the internal elastic response
during the collision.)

\incompP is not a phenomenon, but an ontic property; its empirical results arise
probabilistically in \statP. \incompP is seen to be true for the classical system
presented here by analysis of that system; it is \emph{suspected} to be true of quantum
mechanics from numerous analyses of the experimental conditions necessary to manifest
distinct variables, all of which support this suspicion. (\incompP could be shown to be
\emph{true} of quantum mechanics only through an analysis of its internal workings, but
(as far as we know or believe) quantum mechanics \emph{has} no internal workings.)

Value-indeterminacy (\indetP), the ``non-reality'' of variables, is (obviously) the norm
for the variables of a nondeterministic system: their values leap into existence at each
event. In the case of several variables, it is \emph{repeatability} (per \vN)---the
value-determinacy of an already-observed variable---which makes the value-indeterminacy
of the other variables seem ``wrong.'' Non-reality of this kind is a problem only given a
belief in an underlying determinism for quantum mechanics. Whatever metaphysical
difficulties \allP---incompatibility, nondeterminism and nonrealism---may bring, such
difficulties are not particular to quantum mechanics, nor do they call for heroic
efforts of quantum interpretation; to the extent that an interpretation of quantum
mechanics is based on \emph{these} properties, it is irrelevant.

\bigskip\bigskip
\appendix
\setcounter{equation}{0}%
\section{Elementary probability of propositions}\label{S:ElemProb}%
Each event of a probabilistic system is characterized by the set of propositions which
take on truth values; there can be no presumption that all propositions regarding the
system need take on a truth value at each event.%
\footnote{One might draw a card from a deck, or one might throw the deck into the air;
the proposition $p=\text{``A card landed on the chair''}$ is true or false in the second
event, but has neither meaning nor truth value in the first event.} %
For all propositions which do take on truth values in a given event we have the following:%
\footnote{Disjunction (``or'') is indicated by $\vee$; conjunction (``and''), by
$\wedge$; negation (``not'') by $\Not$.} %
\begin{subequations}
\begin{align}
  &\Pr[]{\text{F}}=0\leq\Pr[]{a}\leq1=\Pr[]{\text{T}} \\
  &\Pr[]{a\vee b}+\Pr[]{a\wedge b}=\Pr[]{a}+\Pr[]{b} ,
\end{align}
with F and T the absurd and trivial propositions, respectively; hence
\begin{equation}
  \Pr[]{\Not a}=1-\Pr[]{a}.
\end{equation}\end{subequations}

The set \set{a_j} is \emph{disjoint} \IFF, whenever all  \set{a_j} take on values,
$\;a_j\wedge a_{j'}=F,\;j\neq j'$; a disjoint set satisfies
$\;\Sum{t}\Pr[s]{a_t}=\Pr[s]{\Orj{t}{a_t}}$ for all preparations $s$. The set \set{a_j}
is \emph{complete} \IFF, whenever all  \set{a_j} take on values,
$\;\Orj{t}{a_t}=\text{T}$; hence, a disjoint, complete set satisfies
$\Sum{t}\Pr[s]{a_t}=1$ for all preparations $s$.

The conditional probability (probability conditioned on an occurrent fact),  defined by
\begin{equation}\label{E:CondProb}
 \Prob[]{b}{a}=
 \begin{cases}
   \Pr[]{a\wedge b}/\Pr[]{a}&\Pr[]{a}>0\\
   \text{undefined}&\text{otherwise},
 \end{cases}
\end{equation}
is the probability of the truth of the proposition $b$ given that the fact stated by the
proposition $a$ occurs. Examples of the condition $a$ would be ``the coin was flipped,''
``the Jokers were removed from the deck,'' ``the first card was a King''; equally well
(though not seen in this paper), the condition may be yet to occur: ``the probability of
drawing a King given that the card drawn after it is a Spade.'' The condition is an
occurrent fact---the only place, in a probabilistic theory, where ``what actually
happens'' can appear.

Given the disjoint and complete set \set{\p{j}} and an arbitrary value $q$, what can be
said of $\Sum{t}\Pr[s]{\p{t}\wedge q}$? Because \set{\p{j}} is disjoint,
\set{\p{j}\wedge q} is disjoint; hence $\Sum{t}\Pr[s]{\p{t}\wedge q}
=\Pr[s]{(\Orj{t}{\p{t}})\wedge q}$. Since the set \set{\p{j}} is complete,
$\Orj{t}{\p{t}}=\text{T}$, thus
\begin{equation}\label{E:MargProb}
 \Sum{t}{\Pr[s]{\p{t}\wedge q}}=\Pr[s]{q},
\end{equation}
the formula of marginal probability. (The use of the term ``marginal'' refers to row- and
column-sums in the margins of a table of probabilities of $\p{j}\wedge q_k$.)

\section{The system of \RefSec{S:Mimick}}%
\label{S:development}%
System: a deck of cards marked with three%
\footnote{The generalization to more than three variables is obvious, and has no effect
on the results, \RefEqns{E:PrXZ}, \eqref{E:PpkandYgX}, \eqref{E:generalIntA}, and
\eqref{E:Interference}. } %
``variables,'' $P$, $Q$, and $R$ (think of \Face, \Suit, and, say, \Letter), each taking
on $V$ values denoted respectively $p_k$, $q_l$, and $r_m$. The specific card denoted
$(p_k\cdot q_l\cdot r_m)$ appears $N(p_k\cdot q_l\cdot r_m)$ times in the deck. We have
(in all permutations) $N(p_k\cdot q_l)=\Sum{m}N(p_k\cdot q_l\cdot r_m)$ and
 $N(p_k)=\Sum{lm}N(p_k\cdot q_l\cdot r_m)$. The restriction that each value of each
variable has equal \emph{a priori} probability requires $N(p_k)=N(q_l)=N(r_m)\DefEq N$;
the total number of cards in the deck is then $NV$, and the fractional occurrence of each
card in the deck is
\begin{equation}
 f(p_k\cdot q_l\cdot r_m)=\frac{n_{klm}}{V},
\end{equation}
where
\begin{equation}
 n_{klm}\equiv\frac{N(p_k\cdot q_l\cdot r_m)}{N}.
\end{equation}
Note that all \emph{double} sums of $n_{klm}$ equal 1, and all \emph{triple} sums equal
$V$.

\subsection{Analysis of the system}\label{SS:derivgeneral}%
According to the rules of system \SysS (and systems \SysSvi and \SysSvd, as well), the
probability of the occurrence of the specific card $(p_k\cdot q_l\cdot r_m)$ at the top
of the subdeck, the system having been prepared in $x$ and the value $P=p_j$ (or
$Q=q_k$) having been observed, is
\begin{subequations}\label{E:general0}
\begin{align}
  &\Prob[x]{(p_k\cdot q_l\cdot r_m)}{p_j}=%
           \KDelta{k}{j}\,n_{jlm}\label{E:general1}\\
  &\Prob[x]{(p_k\cdot q_l\cdot r_m)}{q_j}=\KDelta{l}{j}\,n_{kjm}.\label{E:general2}
\end{align}\end{subequations}

Summing \RefEqns{E:general0} over $l$ and $m$, we have
\begin{subequations}\label{E:PrXZ}
\begin{align}
 &\Prob[x]{p_k}{p_j}=\KDelta{k}{j}\label{E:PrPP}\\
 &\Prob[x]{p_k}{q_j}=\frac{N(p_k\cdot q_j)}{N}.\label{E:PrPQ}
\end{align}\end{subequations}
\RefEqn{E:PrPQ} establishes
\begin{equation}\label{E:sym}
  \Prob[x]{p_k}{q_j}=\Prob[x]{q_j}{p_k}.
\end{equation}

\Prob[x]{y}{p_k} does not depend on the preparation state $x$ (assuming
$\Pr[x]{p_k}\neq0$)---that is, the system's probabilities are Markovian:
\begin{equation}\label{E:PpkandYgX}
  \Pr[x]{p_k\andthen y}=\Prob{y}{p_k}\Pr[x]{p_k}.
\end{equation}
Henceforth in this Appendix we drop the preparation-state subscript from conditional
probabilities.

Consider the ``marginal probability'' summation (where neither of the variables $X$ or
$Y$ is the variable $P$):
\begin{equation}\label{E:margfail}
 \Sum{t}\Pr[x]{p_t\andthen y}=\Sum{t}\Prob{y}{p_t}\Pr[x]{p_t}%
 =\frac{1}{N^2}\Sum{t}N(p_t\cdot x)N(p_t\cdot y).
\end{equation}
However, \Prob{y}{x} is 0 or 1, if $x$ and $y$ are values of the same variable, or
$N(x\cdot y)/N$, if they are values of distinct variables $X$, $Y$. It is obvious, in
the former case, that $\Sum{t}\Pr[x]{p_t\andthen y}\neq\Pr[x]{y}$; numerical examples
easily establish the occurrence of this inequality in the latter case.

\subsection{Analysis of the system---interference}\label{SS:derivint}%
Introduce the additional variable $\Pi$ with values \set{\pi_j}. $\Pi$ is a function of
the variable $P$, $\Pi=f(P)$; the function is defined by $f(\p{1})=\pi_1$,
$f(\p{2})=\pi_1$, $f(\p{j})=\pi_{j-1},\;j>2$. Then
 $N(\pi_1\cdot q_k\cdot r_l)=N(p_1\cdot q_k\cdot r_l)+N(p_2\cdot q_k\cdot r_l)=2N$, and
 $N(\pi_j\cdot q_k\cdot r_l)=N(p_{j+1}\cdot q_k\cdot r_l)=N$ for $j>1$.
The rule for the observation of a variable is unchanged. (In the example of
\RefSec{S:Mimick}, $P$ is \Suit, \p{1} is \H, \p{2} is \D, \p{3} is \S, $\Pi$ is \Color,
$\pi_1$ is \R and $\pi_2$ is \B.)

The probability of the occurrence of a specific card at the top of the subdeck, the
system having been prepared in $x$ and the value $P=p_j$, $Q=q_k$, or $\Pi=\pi_1$ having
been observed, is
\begin{subequations}\label{E:generalInt0}
\begin{align}
 &\Prob{(\pi_1\cdot q_l\cdot r_m)}{p_j}=%
     (\KDelta{1}{j}+\KDelta{2}{j})\,n_{jlm}\label{E:generalInt0a}\\
 &\Prob{(\pi_1\cdot q_l\cdot r_m)}{q_j}=%
      \KDelta{l}{j}\,(n_{1lm}+n_{2lm})\label{E:generalInt0b}\\
 &\Prob{(p_j\cdot q_l\cdot r_m)}{\pi_1}=%
    (\KDelta{j}{1}\,n_{1lm}+\KDelta{j}{2}\,n_{2lm})/2\label{E:generalInt0c}
\end{align}\end{subequations}

Summing \RefEqns{E:generalInt0a} and \eqref{E:generalInt0c} over $l$ and $m$, we find
\begin{subequations}\label{E:generalIntA}
\begin{equation}\label{E:PrPiPInt}
 \Prob{\pi_1}{p_j}=\Prob{p_1}{p_j}+\Prob{p_2}{p_j}=2\Prob{p_j}{\pi_1};
\end{equation}
summing \RefEqn{E:generalInt0b} over $l$ and $m$, and in \RefEqn{E:generalInt0c}
exchanging $j$ and $l$ and then summing over $l$ and $m$, we find
\begin{equation}\label{E:PrPiQInt}
 \Prob{\pi_1}{q_j}=\Prob{p_1}{q_j}+\Prob{p_2}{q_j}=2\Prob{q_j}{\pi_1}.
\end{equation}
\end{subequations}

In the definition of interference, \RefEqn{E:IntDef}, we take $\EvP=\Pi_1$. Then
\RefEqns{E:generalIntA} show that \RefEqn{E:IntCrit1} is satisfied, and
\RefEqn{E:IntCrit2}, the interference, becomes
\begin{equation}
  \Prob{\pi_1\andthen y}{x}-\Prob{(p_1\vee p_2)\andthen y}{x}.
\end{equation}
Because $\pi_1$ completely specifies the p-state, \RefEqn{E:PpkandYgX} generalizes to
\[\Prob{\pi_1\andthen y}{x}=\break\Prob{y}{\pi_1}\Prob{\pi_1}{x}=%
  \Sum{s,t=1}^2\Prob{y}{p_s}\Prob{p_t}{x}\]
(using \RefEqn{E:generalIntA}). Thus the interference is given by
\begin{equation}\label{E:Interference}
 -\Prob{y}{p_1}\Prob{p_2}{x}-\Prob{y}{p_2}\Prob{p_1}{x},
\end{equation}
which does not vanish in general. (This \emph{classical} interference arises from the
off-diagonal terms of a sum, exactly as in the corresponding quantum-mechanical
expression.)

\section{Exercises}\label{S:exercises}%
These back-of-the-envelope exercises introduce the reader to some of the principles
involved in several-variable stochastic systems.
\begin{itemize}
\item[]Given a deck of cards, we may choose to observe the value of either \Face or \Suit;
the rule for doing so is
\begin{itemize}
\item[1.]Draw a card.
\item[2a.]To observe \Face: report the card's \Face value; if $\Face=\K$, return
the card to the deck, otherwise discard it.
\item[2b.]To observe \Suit: report the card's \Suit value; if $\Suit=\H$, return
the card to the deck, otherwise discard it.
\end{itemize}
The deck is \set{\K\S,\,\Q\H}; hence \set{\K,\,\Q} is a complete set of values of \Face,
and \set{\S,\,\H} is a complete set of values of \Suit.
\end{itemize}

\noindent Exercise 1. (\ref{incompP})\ \ Show that it is not always possible to observe
both \Face and \Suit simultaneously. (Hint: suppose the card drawn is $\K\S$.)\smallskip

\noindent Exercise 2. (\ref{nojointP})\ \ Show that observations of \Suit and \Face do
not commute temporally; for example, show that $\Pr{\K^{[1]}\wedge\S^{[2]}}=1/4$, while
$\Pr{\S^{[1]}\wedge\K^{[2]}}\!=0$.  What does this imply regarding the existence of
joint distributions (such as \Pr{\S\wedge\K})?\smallskip

\noindent Exercise 3. (\ref{margP})\ \ Show that \RefEqn{E:MargProbSeqBad} gives an
ambiguous value for $\Pr{\K^{[2]}}$,  by showing that
$\Pr{\big(\K\vee\Q\big)^{[1]}\wedge\K^{[2]}}=3/4$, while
$\Pr{\big(\S\vee\H\big)^{[1]}\wedge\K^{[2]}}=0$. How can this be reconciled with the
apparent fact that $\K\vee\Q=\S\vee\H=\text{T}$?
\smallskip

(No exercise regarding \nosharpP is possible: because observations on this system do not
repeat, ``sharpness'' has no meaning.)

 \renewcommand{\refname}{\sc References}
 \footnotesize%


\begin{thebibliography}{10}

\bibitem{Jammer74}
M. Jammer, \emph{The Philosophy of Quantum Mechanics}, Wiley, New York, 1974.

\bibitem{Margenau63a}
H. Margenau, ``Measurements in quantum mechanics,'' Ann. Phys. \textbf{23},
  469--485 (1963).

\bibitem{Farina93}
J.~E.~G. Farina, ``An elementary approach to quantum probability,'' Am. J.
  Phys. \textbf{61}(5), 466--468 (1993).

\bibitem{FeynmanVolIII}
R.~P. Feynman, R.~B. Leighton, and M. Sands, \emph{The {F}eynman Lectures on
  Physics, {V}ol. {III}}, Addison-Wesley, 1965.

\bibitem{AlbertAD85}
D.~Z. Albert, Y. Aharonov, and S. D'Amato, ``Curious new statistical prediction
  of quantum mechanics,'' Phys. Rev. Lett. \textbf{54}(1), 5--7 (1985).

\bibitem{Kirkpatrick:ThreeBox}
K.~A. Kirkpatrick, ``Classical {T}hree-{B}ox `paradox','' J. Phys. A,
  quant-ph/0207124.

\bibitem{Bell66}
J.~S. Bell, ``On the problem of hidden variables in quantum mechanics,'' Rev.
  Mod. Phys. \textbf{38}(3), 447--452 (1966).

\bibitem{KochenSpecker67}
S. Kochen and E.~P. Specker, ``The problem of hidden variables in quantum
  mechanics,'' J. Math. Mech. \textbf{17}(1), 59--87 (1967).

\bibitem{Redhead89}
M. Redhead, \emph{Incompleteness, Nonlocality, and Realism}, Clarendon
  Paperbacks, Oxford, 1989.

\bibitem{dEspagnat95}
B. {d}'Espagnat, \emph{Veiled Reality}, Addison-Wesley, 1995.

\bibitem{Fine82b}
A. Fine, ``Joint distributions, quantum correlations, and commuting
  observables,'' J. Math. Phys. \textbf{23}(7), 1306--1310 (1982).

\end{thebibliography}
\end{document}